 \documentclass[prb,twocolumn,showpacs,preprintnumbers,amsmath,amssymb]{revtex4}

\usepackage[dvips]{graphicx}% Include figure files
\usepackage{dcolumn}% Align table columns on decimal point
\usepackage{bm}% bold math

\begin{document}

%\preprint{APS/123-QED}

\title{  
Electron Transport Through Molecules: \\
Self-consistent and Non-self-consistent Approaches 
}

\author{ San-Huang Ke,$^{1,2}$ Harold U. Baranger,$^{2}$ and Weitao Yang$^{1}$}

\affiliation{
$^{\rm 1}$Department of Chemistry, Duke University, Durham, NC 27708-0354 \\
$^{\rm 2}$Department of Physics, Duke University, Durham, NC 27708-0305
}

\date{\today}% It is always \today, today,

\begin{abstract}

A self-consistent method for calculating electron transport through a molecular 
device is developed. It is based on density functional theory electronic structure 
calculations under periodic boundary conditions and implemented in the framework 
of the non-equilibrium Green function approach. To avoid the substantial computational 
cost in finding the $I$-$V$ characteristic of large systems, we also develop an approximate 
but much more efficient non-self-consistent method. Here  the change in effective potential 
in the device region caused by a bias is approximated by the main features of the voltage drop.  
As applications, the $I$-$V$ curves of a carbon chain and an aluminum chain sandwiched between 
two aluminum electrodes are calculated -- two systems in which the voltage drops very differently.
By comparing to the self-consistent results, we show that this non-self-consistent approach 
works well and can give quantitatively good results.         

\end{abstract}

\maketitle

%===================================================================
\section{Introduction}

In recent years, electron transport through molecules sandwiched between metallic electrodes 
has been attracting increasing attention both for fundamental reasons and because it may form 
the basis of a future molecular electronics technology.\cite{mol1,mol2,mol3,mol4,mol5}  
Experimentally, it is difficult to precisely manipulate or even measure the atomic structure 
of the molecule-electrode contacts. Therefore, neither the influence of atomic structure on 
transport through the devices nor a path to improved performance is clear.  As a result, 
the ability to calculate the atomic and electronic structure as well as the transport 
properties of electrode-molecule-electrode systems is important and useful in this field.  

Electron transport through nanoscale molecular devices differs significantly from that through 
macroscopic semiconductor heterostructures. In the latter, the effective-mass approximation 
is generally successful because of the periodic lattice structure and large electron wavelength.  
In contrast, in a molecular device a carrier electron will be scattered by only a few atoms whose 
particular arrangement, then, matters a great deal. Consequently, the effective-mass approximation
breaks down, and the electronic structure of the molecular device must be taken into account 
explicitly. For this purpose, methods based on density functional theory (DFT) are sufficiently 
accurate and efficient.\cite{dft1,dft2}  Conventional DFT methods, however, deal with either 
closed molecular systems (in quantum chemistry) or periodic solids (in solid state physics), 
neither of which is applicable to molecular transport. Thus one needs to develop a DFT approach 
suitable for a system which is open, infinite, non-periodic, and non-equilibrium (if the bias 
voltage is nonzero).

One way to do this was suggested by Lang, {\it et al.} \cite{lang1,lang2,lang3,lang4} 
By using the jellium model for the two metallic electrodes of an electrode-molecule-electrode 
system, they  mapped the Kohn-Sham equation of the system into the Lippmann-Schwinger scattering 
equation and solved for the scattering states self-consistently.  They then calculated the 
current by summing up the contributions from all the scattering states, following a 
Landauer-type approach.\cite{landauer1}  In this way, both the conductance and $I$-$V$ 
characteristics of the system can be investigated.  The use of the jellium model for electrodes 
is convenient and simple but limited: it cannot include the effects of different contact 
geometries and surface relaxation, for instance.  It also cannot deal with directional bonding 
such as in semiconductors and transition metals. As a result, the molecule-electrode charge 
transfer, which is one of the key factors affecting transport, may not be quantitatively 
correct.\cite{test}

Another way to develop the desired DFT approach is to use the non-equilibrium Green function 
(NEGF) method.\cite{negf1,negf2} The required open and non-equilibrium conditions can be 
treated rigorously, at least in a formal sense. This method is also closely related to the 
Landaur approach\cite{landauer1} and has proven to be powerful for studying electron transport 
through nanoscale devices.  Therefore, by combining the NEGF method with conventional DFT-based 
electronic structure methods used in quantum chemistry or solid-state physics, the coherent 
transport properties of an electrode-molecule-electrode system can be determined fully 
self-consistently from first-principles.  A further advantage of the NEGF+DFT combination is 
that the atomic structure of the device region and the metallic electrodes are treated 
explicitly on the same footing.  As has been mentioned, the molecule-electrode interaction 
will induce charge transfer between them and atomic relaxation of their contact -- both have 
a significant effect on electron transport. As a result, the division of the system into the 
molecule and the electrodes is not meaningful anymore, and some parts of the electrodes must 
be included into the device region to form an ``extended molecule".

Based on this combined NEGF+DFT method, there are several successful implementations 
\cite{datta1, datta2, transiesta, mcdcal,mono} for molecular conduction and extensive
theoretical results in the recent literature.
\cite{datta1, datta2, transiesta, mcdcal, xue, transiesta1, mcdcal1,mono} 
According to the way of treating the extended molecule, the semi-infinite leads, 
and their couplings in a lead-molecule-lead (LML) system,
these implementations can be roughly divided into two categories.

In one category people adopted a cluster geometry
for all the subsystems of a LML system 
or for the extended molecule with the leads
treated by a tight binding approach (for example, Ref. \onlinecite{datta1,xue,non-sc}).
It is then convenient to
employ well-established quantum chemistry code (like Gaussian or DMol) to do
the electronic structure calculation for the subsystem(s).
However, there are potential problems with these treatments for strong
molecule-lead couplings:
in this case it is obviously necessary to included large parts of the
leads into the extended molecule so that the strong molecule-lead interaction
can be fully accommodated. To eliminate the artificially introduced surface
effects an even larger system is needed, which is usually difficult
to deal with by a quantum chemistry code. So in practice only several (or
even only one) lead atoms are attached to the molecule to form an extended
molecule (for example, Ref. \onlinecite{xue,non-sc}).
In this case, significant artificial surface effects are inevitable, the contact
atomic relaxation cannot be included, and an accurate molecule-lead
coupling is not available. In addition, there may be artificial scattering at the interface
between the tight-binding part of the lead and the DFT part of the lead
(included in the extended molecule). 

In the other category (Ref.\onlinecite{transiesta,mcdcal}), 
people adopted periodic boundary conditions (PBC)
(as in solid state physics) with large parts of the leads included
in the extended molecule, so that the interaction between the molecule and its
images will be screened off by the metallic lead in between. In this case
all the potential problems mentioned above will be absent and the 
whole LML system becomes nearly perfect in geometry and all the subsystems are 
treated exactly on the same footing.
Two examples of successful implementations adopting the PBC are 
the TranSiesta package \cite{transiesta} and the MCDCAL package.\cite{mcdcal}
On the other hand, a drawback is introduced by PBC: when a bias is applied,
the Hartree potential must jump unphysically between unit cells. This has previously been
addressed by having an independent solution of the Poisson equation. 

In this paper, we first develop a fully self-consistent NEGF+DFT method with
PBC, which has small but important differences from the two previous implementations. 
The advantage of our 
method is that it is simple  while still rigorous: the non-equilibrium condition under a bias 
is fully included in the NEGF part, and, as a result, we do not need to make changes in the 
conventional electronic structure part. So it is straightforward to combine with any electronic 
structure method that uses a localized basis set.  More importantly, in this way the problem 
of the unphysical jumps in the Hartree potential is avoided.

A shortcoming of the full self-consistent (SC) NEGF+DFT approach is the large computational 
effort involved, especially for large systems, large bias voltages, or cases where many bias 
voltages need to be calculated as for $I$-$V$ characteristics.  As a result, a 
non-self-consistent (non-SC) method with much higher efficiency and useful accuracy 
is highly desirable.
As a step toward this goal, we also construct 
an approximate but much more efficient non-SC method in which the change in self-consistent 
effective potential in the device region caused by a bias is approximated by the main features 
of the voltage drop.

As an application of our approach, in this paper we do calculations by combining it with a very efficient electronic structure package SIESTA.\cite{siesta}  The $I$-$V$ curves of two systems with different typical voltage drop behaviors -- a carbon or aluminum chain sandwiched between two aluminum electrodes -- are calculated. Our self-consistent results are in good agreement with those from previous calculations.\cite{transiesta,mcdcal_chain}
By comparison to the self-consistent results, we examine the validity of the non-SC approach, showing that this approach works quite well and can give quantitatively nearly correct answers.

The arrangement of this paper is as follows. In Section II we give briefly a description of our implementation of the NEGF+DFT method.  Because the basic formalism of the NEGF+DFT is well established,\cite{datta1, datta2, transiesta, mcdcal} we show only those formula useful for introducing the new features of our method.  The present SC and non-SC approaches are explained in Section III.  Section IV starts with results for a carbon chain and an aluminum chain sandwiched between two Al(001) electrodes.  Our results are compared with previous results, and we discuss the validity of the non-SC approach by comparison to the self-consistent results. In Section V we summarize and conclude.

\section{NEGF+DFT method and its implementation}

\subsection{Modeling of real physical systems}

%------------------------------ Fig.1 ---------------------------------
\begin{figure}[b] 
\includegraphics[angle=270,width= 8.5cm]{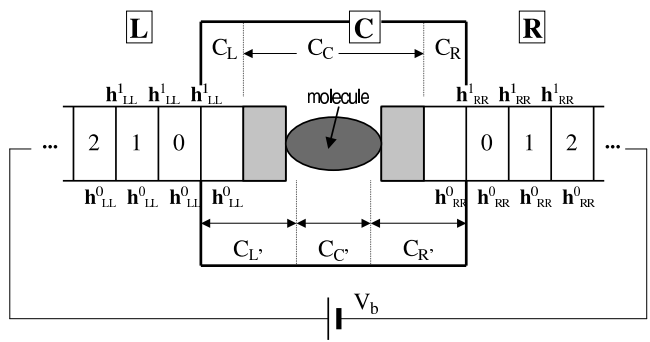}
\label{fig_str}
\caption{Schematic drawing of a system containing a molecule sandwiched between two metallic electrodes (leads $L$ and $R$).  The region $C$ is formed by including some parts of $L$ and $R$ so that the $C_C$ part (extended molecule) is charge neutral and the $C_L$ and $C_R$ parts have bulk properties. Because of the use of a localized basis set, the leads $L$ and $R$ can be divided into principal layers (denoted by numbers 0, 1, 2, ...). $C_{L'}$, $C_{C'}$, and $C_{R'}$ denote the parts used in the present non-SC approach (see Section II E). Their interface is called $X$ in the text.  $\mathbf{h}^{0}$'s and $\mathbf{h}^{1}$'s are the Hamiltonian matrices within and between the principal layers, respectively. }
\end{figure}

Experimentally, a molecular device system consists of at least a molecular junction coupled with two metallic electrodes (leads $L$ and $R$) under a bias $V_b$ (two-terminal system). In some cases, there is also a gate terminal applying a gate voltage on the whole system (three-terminal system). Here we consider only the two-terminal system which is schematically shown in Fig. 1. An important consideration for modeling the real physical system is the charge transfer and atomic relaxation around the two molecule-lead contact regions.  As a result, we have to include some parts of the metallic leads into the device region, forming an extended molecule. One obvious convergence criteria for the size of the extended molecule is its charge neutrality. Then the charge transfer and the potential disturbance caused by the molecule can be considered screened off outside the extended molecule region.  In order to obtain good convergence, we actually include a large part of each metallic lead into the $C$ region, so that the layers adjacent to $L$ and $R$ (i.e., $C_L$ and $C_R$ parts in Fig. 1) have bulk properties. The total Hamiltonian of the system is:
\begin{equation}
\hat{H} = \hat{H}_{LL} + \hat{H}_{CC} + \hat{H}_{RR} + \hat{H}_{LC} + \hat{H}_{CR}.
\end{equation}
Note that, here the leads $L$ and $R$ interact only through the molecular junction, so their direct interaction term $\hat{H}_{LR}$ vanishes (this can always be satisfied by using a localized basis set).

\subsection{Localized basis set}
When $\hat{H}$ is expanded in a basis set, generally only the matrix of $\hat{H}_{CC}$ (denoted $\mathbf{H}_{CC}$) is finite.  However, consider a localized (but not necessarily orthogonal) basis set, by which we mean that the overlap between any two basis functions, $\phi_{\mu}({\mathbf{r}-\mathbf{R}_1})$ and $\phi_{\nu}({\mathbf{r}-\mathbf{R}_2})$, will be zero if they are separated far enough from each other: $\mathbf{S}_{\mu \nu} \equiv \langle \mu|\nu \rangle =0$ if $|{\mathbf{R}_1-\mathbf{R}_2}| >$ certain cutoff distance. In this case, the region $C$ interacts directly only with finite parts of $L$ and $R$, and the non-zero part of the matrices $\mathbf{H}_{LC}$ and $\mathbf{H}_{CR}$ also become finite.  Furthermore, we can divide the leads $L$ and $R$ into principal layers so that any principal layer interacts only with its two nearest neighbors (see Fig. 1).  As a result, the matrices $\mathbf{H}_{LL}$ and $\mathbf{H}_{RR}$ have the following block tridiagonal form:
\begin{eqnarray} \label{equ_HLL}
(\mathbf{H}_{LL})_{ij}&=&
\left\{ \begin{array}{ll}
 \mathbf{h}_{LL}^{0}, & \textrm{if } i-j =0\\*[0.04in]
 \mathbf{h}_{LL}^{1}, & \textrm{if } i-j =1\\*[0.04in]
(\mathbf{h}_{LL}^{1})^{\dagger}, & \textrm{if } j-i =1\\*[0.04in]
 0, & \textrm{if } |i-j|>1, 
\end{array} \right. 
%\\
%(\mathbf{H}_{RR})_{ij}&=&
%\left\{ \begin{array}{ll}
% \mathbf{h}_{RR}^{0}, & \textrm{if } i-j =0\\
% \mathbf{h}_{RR}^{1}, & \textrm{if } j-i =1\\
%(\mathbf{h}_{LL}^{1})^{\dagger}, & \textrm{if } i-j =1\\
% 0, & \textrm{if } |i-j|>1, 
%\end{array} \right.
\end{eqnarray}
where $\mathbf{h}^{0}_{LL}$ and $\mathbf{h}^{1}_{LL}$ are the Hamiltonian matrices within and between the principal layers, respectively, and $i$, $j$ are principal layer indexes as shown in Fig. 1.  Because the $C_L$ and $C_R$ parts which interact directly with $L$ and $R$, have bulk properties, the non-zero part of $\mathbf{H}_{LC}$ ($\mathbf{H}_{CR}$) is just $\mathbf{h}_{LL}^1$ ($\mathbf{h}_{RR}^1$), as shown in Fig. 1.

In the localized basis and after the partition shown in Fig. 1, the matrix Green function $\mathbf{G}$ of the whole system, defined by
%\begin{equation}
$(E \mathbf{S} - \mathbf{H}) \mathbf{G}(E) = \mathbf{I}$,
%\end{equation}
satisfies
\begin{eqnarray}
\left[ \begin{array}{ccc}
E\mathbf{S}_{LL}-\mathbf{H}_{LL} & E\mathbf{S}_{LC}-\mathbf{H}_{LC} & 0 \\[0.04in]
E\mathbf{S}_{LC}^{\dagger}-\mathbf{H}_{LC}^{\dagger} & E\mathbf{S}_{CC}-\mathbf{H}_{CC} & 
                                                       E\mathbf{S}_{CR}-\mathbf{H}_{CR} \\[0.04in]
0 & E\mathbf{S}_{CR}^{\dagger}-\mathbf{H}_{CR}^{\dagger} & E\mathbf{S}_{RR}-\mathbf{H}_{RR}
\end{array} \right] \\ \nonumber 
\times 
\left[ \begin{array}{ccc}
\mathbf{G}_{LL} & \mathbf{G}_{LC} & \mathbf{G}_{LR} \\
\mathbf{G}_{CL} & \mathbf{G}_{CC} & \mathbf{G}_{CR} \\
\mathbf{G}_{RL} & \mathbf{G}_{RC} & \mathbf{G}_{RR}
\end{array} \right]
=
\left[ \begin{array}{ccc}
\mathbf{I}_{LL} & 0 & 0 \\
0 & \mathbf{I}_{CC} & 0 \\
0 & 0 & \mathbf{I}_{RR}
\end{array} \right].
\end{eqnarray}
The most important part of $\mathbf{G}$ is $\mathbf{G}_{CC}$, corresponding to the region $C$; from the above equation,
\begin{equation}
\label{equ_Gc}
\mathbf{G}_{CC}(E)=\left\{E\mathbf{S}_{CC}-\left[\mathbf{H}_{CC}+
                \mathbf{\Sigma}_L(E)+\mathbf{\Sigma}_R(E)\right]\right\}^{-1},
\end{equation}
where $\mathbf{\Sigma}_L(E)$ and $\mathbf{\Sigma}_R(E)$ are self-energies which incorporate the effect of the two semi-infinite leads $L$ and $R$, respectively.
$\mathbf{\Sigma}_L(E)$, for example, is defined by
\begin{eqnarray}
\mathbf{\Sigma}_L(E)\!=\!\left( E\mathbf{S}_{LC}-\mathbf{H}_{LC}\right)^{\dagger}
                     \mathbf{G}_{LL}^0(E)\left( E\mathbf{S}_{LC}-\mathbf{H}_{LC}\right)
%\\
%\mathbf{\Sigma}_R(E)\!=\!\left( E\mathbf{S}_{CR}-\mathbf{H}_{CR}\right)
%                     \mathbf{G}_{RR}^0(E)\left( E\mathbf{S}_{CR}-\mathbf{H}_{CR}\right)^{\dagger},
\end{eqnarray}
where $\mathbf{G}_{LL}^0$ 
%and $\mathbf{G}_{RR}^0$ 
is the retarded Green function of the left semi-infinite lead. The latter is given in turn by
\begin{eqnarray}
\mathbf{G}_{LL}^0(E)&=&\left(z\mathbf{S}_{LL}-\mathbf{H}_{LL}\right)^{-1},
\\
%\mathbf{G}_{RR}^0(E)&=&\left(z\mathbf{S}_{RR}-\mathbf{H}_{RR}\right)^{-1},\\
z&=&E+i\eta, \nonumber
\end{eqnarray}
where a typical value for the lifetime broadening $\eta$ is about 1~meV.  

Because of the localized basis set, the non-zero part of $\mathbf{S}_{LC}$, $\mathbf{H}_{LC}$, $\mathbf{S}_{CR}$, and $\mathbf{H}_{CR}$ become finite (being $\mathbf{s}_{LL}^1$, $\mathbf{h}_{LL}^1$, $\mathbf{s}_{RR}^1$, and $\mathbf{h}_{RR}^1$).  As a result, only the part of $\mathbf{G}_{LL}^0$ and $\mathbf{G}_{RR}^0$ corresponding to the $0^{th}$ principal layer of the two leads (denoted $\mathbf{g}_{LL}^0$ and $\mathbf{g}_{RR}^0$) are needed for calculating the non-zero part of the self-energies: 
\begin{eqnarray}
\mathbf{\Sigma}_L(E)\!=\!\left( E\mathbf{s}_{LL}^1-\mathbf{h}_{LL}^1\right)^{\dagger}
                    \mathbf{g}_{LL}^0(E)\left( E\mathbf{s}_{LL}^1-\mathbf{h}_{LL}^1\right). 
%\\
%\mathbf{\Sigma}_R(E)\!=\!\left( E\mathbf{s}^1_{RR}-\mathbf{h}_{RR}^1\right)
%        \mathbf{g}_{RR}^0(E)\left( E\mathbf{s}_{RR}^1-\mathbf{h}_{RR}^1\right)^{\dagger}.
\end{eqnarray}
Our notation here follows that for the Hamiltonian: $\mathbf{s}$ and $\mathbf{g}$ are submatrices of the corresponding upper case matrices.  $\mathbf{g}_{LL}^0$ and $\mathbf{g}_{RR}^0$ are simply the surface Green functions of the two semi-infinite leads.  $\mathbf{g}_{LL}^0$, for example, can be calculated either by simple block recursion,
%\begin{widetext}
\begin{eqnarray}
\mathbf{g}_{LL}^0(E)&=&\big[ z\mathbf{s}_{LL}^0 - \mathbf{h}_{LL}^0 \\ 
 &-& \left( z\mathbf{s}_{LL}^1-\mathbf{h}_{LL}^1\right)^{\dagger} 
 \mathbf{g}_{LL}^0(E)\left( z\mathbf{s}_{LL}^1-\mathbf{h}_{LL}^1\right) \big]^{-1}, \nonumber
%\\
%\mathbf{g}_{RR}^0(E)&=&\left[ z\mathbf{s}_{RR}^0 - \mathbf{h}_{RR}^0 
% - \left( z\mathbf{s}_{RR}^1-\mathbf{h}_{RR}^1\right) 
% \mathbf{g}_{RR}^0(E)\left( z\mathbf{s}_{RR}^1-\mathbf{h}_{RR}^1\right)^{\dagger} \right]^{-1},
\end{eqnarray}
%\end{widetext}
or by a renormalization method \cite{renorm} in terms of 
$\mathbf{s}_{LL}^0$, $\mathbf{s}_{LL}^1$, $\mathbf{h}_{LL}^0$, and $\mathbf{h}_{LL}^1$
which can be determined by separate DFT calculations for the two leads.  For small lifetime broadening $\eta$ (1~meV), we find that the renormalization method is much faster than simple block recursion. This is natural since $n$ renormalization interations incorporate $2^n$ principal layers, while $n$ recursions incorporate only $n$. 

 From $\mathbf{G}_{CC}(E)$, the projected density of states (PDOS) on the molecule (indicated by $m$) is given by
\begin{equation}
N_{m}(E)=-\frac{1}{\pi} \textrm{Im} \left\{ \textrm{Tr}_{m} 
\left[\mathbf{G}_{CC}(E+i\eta)\cdot \mathbf{S}_{CC}\right]\right\},
\end{equation}
where $\textrm{Tr}_{m}$ means the trace is performed only on the molecular part of the matrix.

%\subsection{NEGF+DFT techniques}
\subsection{Current}

The Non-Equilibrium Green Function technique (NEGF)\cite{negf2,negf_i, negf2, negf_i2} provides a convenient way to calculate the current by post-processing a DFT calculation. The result is quite natural and intuitive: First, the basic assumption is that there is no energy relaxation within the molecular region. Then, following a Landauer-like point of view,\cite{landauer1,negf2} one divides the electrons in the molecule into two sets using scattering-wave states, those that came from the left lead and those that came from the right. The left-lead states are, of course, filled up to the chemical potential in the left lead, $\mu_L$, while the right-lead states are filled up to $\mu_R$. In equilibrium, the two chemical potentials are equal, and the current carried by the left-lead states is, of course, equal to that carried by the right-lead states. As a bias is applied, the balance between the two types of states is disrupted and current flows. As different states are populated because of the change in chemical potentials, the charge density in the molecule also changes. The potential profile must be solved for self-consistently in order to get an accurate measure of the transmission. It is this self-consistency which is the time-consuming part of the calculation.

The expression for the steady-state current through the $C$ region for applied bias $V_b$ is
   \begin{equation} \label{equ_I}
I(V_b)= - \frac{2e^2}{h}\int^{+\infty}_{-\infty}\!\!\!\!T(E,V_b)
         \left[f(E\!-\!\mu_L)-f(E\!-\!\mu_R)\right]dE,
   \end{equation}
where $\mu_L$ and $\mu_R$ are the chemical potentials, $f$ is the Fermi function, and $T(E,V_b)$ is the transmission probability for electrons from the left lead to right lead with energy $E$ under bias $V_b$. The transmission probability is related to Green functions by
   \begin{equation} \label{equ_t}
 T(E,V_b)=\textrm{Tr}\left[\mathbf{\Gamma}_L(E)\mathbf{G}_{CC}(E)
         \mathbf{\Gamma}_R(E)\mathbf{G}_{CC}^{\dagger}(E)\right],
   \end{equation}
where
   \begin{eqnarray} 
 \mathbf{\Gamma}_{L,R}(E) &=&i\left(\mathbf{\Sigma}_{L,R}(E)                        
      -\left[\mathbf{\Sigma}_{L,R}(E)\right]^{\dagger}\right)
   \end{eqnarray} 
reflect the coupling at energy $E$ between the $C$ region and the leads $L$ and $R$, respectively.

The charge density corresponding to the above picture of left-lead states filled to $\mu_L$ and right-lead states filled $\mu_R$ can also be expressed in terms of Green functions. In particular, the density matrix of region $C$ in the basis-function space is
\begin{widetext}
\begin{eqnarray}
\mathbf{D}_{CC}&=&\frac{1}{2\pi}\int_{-\infty}^{+\infty}dE\left[ \label{equ_d1} \label{equ_d2} 
         \mathbf{G}_{CC}(E)\mathbf{\Gamma}_L(E)\mathbf{G}_{CC}^{\dagger}(E)f(E-\mu_L) 
       + \mathbf{G}_{CC}(E)\mathbf{\Gamma}_R(E)\mathbf{G}_{CC}^{\dagger}(E)f(E-\mu_R)\right] \\ 
      &=&-\frac{1}{\pi}\int_{-\infty}^{+\infty}dE\textrm{Im} \label{equ_d3}
           \left[\mathbf{G}_{CC}(E)f(E-\mu_L)\right] \\ \nonumber   
     &&+ \frac{1}{2\pi}\int_{-\infty}^{+\infty}dE\left[
           \mathbf{G}_{CC}(E)\mathbf{\Gamma}_R(E)\mathbf{G}_{CC}^{\dagger}(E)\right] 
       \times  \left[f(E-\mu_R)-f(E-\mu_L) \right].       
\end{eqnarray} 
\end{widetext}
Time-reversal symmetry ($\mathbf{G}_{CC}^{\dagger}=\mathbf{G}_{CC}^{*}$) was invoked in going from (\ref{equ_d2}) to (\ref{equ_d3}).  The integrand of the first term of (\ref{equ_d3}) is analytic (all poles of $\mathbf{G}_{CC}(E)$ are on real axis), so the integral can be evaluated easily by complex contour integration.  However, the integrand of the second term is not analytic, so it must be evaluated by integrating very close to the real axis using a very fine energy mesh. The whole integration path\cite{datta2} is shown in Fig. 2. Because we construct the region $C$ such that $C_L$ and $C_R$ have essentially bulk properties, we can use the bulk density matrix for them.

%------------------------------------ Fig.2 ----------------------------
\begin{figure}[b] \label{fig_path}
\includegraphics[angle=0,width= 5.0cm]{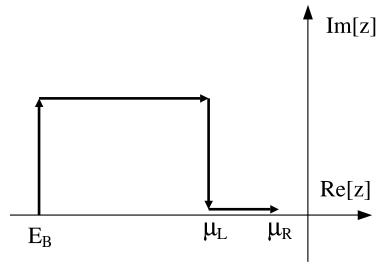}
\caption{Schematic drawing of the integration path in the complex energy plane used to calculate the density matrix [Eq. (\ref{equ_d3})].  $E_B$ is the lowest energy of occupied states, and $\mu_{L,R}$ are the chemical potentials of the left and right leads, respectively ($\mu_{L}$ $<$ $\mu_{R}$ is assumed).  Note that for energy window [$E_B$,$\mu_{L}$] a complex contour integration is performed while for energy window [$\mu_{L}$,$\mu_{R}$] a direct energy integration is performed by using a fine energy mesh and a very small imaginary part.} 
\end{figure}

The calculated density matrix is then output to the DFT part to calculate the electron density $\rho$ and to construct a new $\mathbf{H}_{CC}$:
\begin{equation}
\rho(\mathbf{r})=\sum_{\mu,\nu}\phi^{*}_{\mu}(\mathbf{r})\textrm{Re}
                  \left[\left(\mathbf{D}_{CC}\right)_{\mu\nu}\right]\phi_{\nu}(\mathbf{r}),
\end{equation}
\vspace*{-5mm}
\begin{equation}
(\mathbf{H}_{CC})_{\mu\nu}=\left\langle\mu\left|\hat{T}+\hat{V}_{\textrm{ext}}(\mathbf{r})
                 +\hat{V}_{\textrm{H}}[\rho(\mathbf{r})]
                 +\hat{V}_{\textrm{xc}}[\rho(\mathbf{r})]\right|\nu\right\rangle, \label{equ_H}
\end{equation}
where $\hat{T}$ is the kinetic energy, and $\hat{V}_{\textrm{ext}}$, $\hat{V}_{\textrm{H}}$, and $\hat{V}_{\textrm{xc}}$ are the external, Hartree, and exchange-correlation potential energies, respectively.  The new $\mathbf{H}_{CC}$ replaces the old, a new $\mathbf{D}_{CC}$ is calculated, and so on until $\mathbf{H}_{CC}$ or $\mathbf{D}_{CC}$ converges.  Finally, the transmission function $T(E)$ can be calculated by Eq. (\ref{equ_t}).

One subtlety here is the different boundary conditions used in the Green function and DFT parts 
-- open versus periodic, respectively. 
This means that some iteration must be done even at $V_b = 0$. 
If the supercell of the DFT part has the same size as region $C$ in Fig. 1, 
then the $C_L$ and $C_R$ parts will interact directly due to the periodic boundary condition. 
However, this interaction is absent in the calculation of the density matrix. 
The same problem exists also for $\mathbf{H}_{CC}$.  
So we need to do some translation work between the Green function and DFT parts: 
to add this interaction when we go from NEGF to DFT by using the density matrix elements 
between two adjacent principal layers, and to remove it when we go from DFT to NEGF 
by setting corresponding parts of $\mathbf{S}_{CC}$ and $\mathbf{H}_{CC}$ to zero. 
Generally, the supercell of the DFT part can be made larger than the size of the region $C$,
especially for systems without a translational symmetry, because the DFT part is usually 
much cheaper than the Green function part.

\section{New self-consistent and non-self-consistent approaches}

\subsection{Bias voltage}

For non-zero $V_b$, care must be taken to account for the effects of the bias voltage on the charge density. One way to proceed is to apply a constant field in the direction parallel to the leads within the supercell of the DFT calculation. Thus a linear drop is added to the external potential in Eq. (\ref{equ_H}), and the effect on the charge density follows from, for instance, solving the Poisson equation. This approach is not straightforward for periodic boundary conditions because of the artificial potential jumps at the two supercell boundaries in the lead direction.  One way to eliminate the unphysical jumps is to use a larger supercell for the DFT calculation and replace the Hamiltonian of each of the regions near the potential jumps by the bulk one with a constant potential shift given by the bias voltage; this is implemented in the Transiesta program.\cite{transiesta}

Here we propose a different approach to handle the bias, one which is less obvious but turns out to be simpler in the end: The bias is included through the density matrix ($\mathbf{D}_{CC}$) in the Green function calculation instead of the potential ($\mathbf{H}_{CC}$) in the DFT part.  Specifically, we calculate the density matrix by Eq. (\ref{equ_d1}) under the boundary condition that there is a potential difference $V_b$ between part $C_L$ (together with the left lead) and part $C_R$ (together with the right lead).  This is done by shifting all the potentials related to the left (right) lead and the $C_L$ ($C_R$) part by $-V_b/2$ ($+V_b/2$).  Shifting the potential in a lead is equivalent to directly shifting the energy by the opposite amount, so
\begin{widetext}
\begin{equation}
\mathbf{D}_{CC} = \frac{1}{2\pi}\int_{-\infty}^{+\infty}dE\left[ \label{equ_d4}
            \mathbf{G}_{CC}(E)\mathbf{\Gamma}_L(E+\frac{eV_b}{2})
            \mathbf{G}_{CC}^{\dagger}(E)f(E-\mu_L) 
           + \mathbf{G}_{CC}(E)\mathbf{\Gamma}_R(E-\frac{eV_b}{2})
            \mathbf{G}_{CC}^{\dagger}(E)f(E-\mu_R)\right] \,.
\end{equation}
Here the Green function $\mathbf{G}_{CC}(E)$ has all the potential shifts included,
\begin{equation}
\mathbf{G}_{CC}(E) = \left(E\mathbf{S}_{CC}-\left[\mathbf{H}_{CC}
 +\Delta\mathbf{H}_{C_L}+\Delta\mathbf{H}_{C_R} 
 +\mathbf{\Sigma}_L(E+\frac{eV_b}{2})
    +\mathbf{\Sigma}_R(E-\frac{eV_b}{2})\right]\right)^{-1}, \label{GccVb}
\end{equation}
\end{widetext}
where the $C_L$ and $C_R$ parts of $\mathbf{H}_{CC}$ are replaced by $\mathbf{h}^0_{LL}$ and $\mathbf{h}^0_{RR}$, and their potential shifts are 
\begin{eqnarray}
\left[\Delta\mathbf{H}_{C_L}\right]_{\mu\nu}&=&
 -\frac{eV_b}{2}
  \left[\mathbf{S}_{CC}\right]_{\mu\nu}, \label{equ_sh1}
  \forall  \mu,\nu \in C_L,\\ 
\left[\Delta\mathbf{H}_{C_R}\right]_{\mu\nu}&=&
 +\frac{eV_b}{2}
  \left[\mathbf{S}_{CC}\right]_{\mu\nu},
  \forall  \mu,\nu \in C_R \,. \label{equ_sh2}
\end{eqnarray}
The actual computational process is exactly the same as that of (\ref{equ_d3}) and Fig. 2, and self-consistency is achieved in the same way as for zero bias. Finally, in the calculation of the transmission, the potential shift present in the self-energies appears in $\Gamma_{L,R}$ as in (\ref{equ_d4}),
\begin{equation} \label{equ_t'}
T(E,V_b)=\textrm{Tr}\Big\{
   \mathbf{\Gamma}_L(E+\frac{eV_b}{2})\mathbf{G}_{CC}(E)
 \mathbf{\Gamma}_R(E-\frac{eV_b}{2})\mathbf{G}_{CC}^{\dagger}(E)\Big\}.
 \end{equation}

While the two approaches mentioned here -- the linear external potential and the potential shift in the leads -- give quite different results in the unphysical non-interacting limit, self-consistency ensures that they give the same result in physical cases. Our approach has the distinct technical advantage that the Green function and DFT calculations are completely disconnected, allowing the transport module to be easily combined with a wide variety of electronic structure codes. 

\subsection{Approximate non-self-consistent approaches}

For large systems under large bias, the full SC approach described above is computationally very difficult. The longest part of a one-time calculation is finding the surface Green functions (for all the points in the energy mesh), even with the fast renormalization method.  However, one only needs to do these calculations once and save the results. The major cost for a full self-consistent calculation is from finding $\mathbf{G}_{CC}(E)$ by Eq. (\ref{equ_Gc}) at the many energies needed for the density matrix, especially for the very fine mesh in the energy window [$\mu_{L}$, $\mu_{R}$] (see Fig. 2).

To avoid the complex procedure and large computational effort of full self-consistency, here we propose an approximate non-SC approach in which the bias is included by (a) applying a potential shift $-eV_b/2$ ($+eV_b/2$) to the left (right) lead through the self-energies as in Eq. (\ref{equ_t'}),
and (b) introducing potential shifts non-self-consistently into the region $C$. The main consideration is that it may be a good approximation to replace the change in self-consistent effective potential caused by a bias by the main features of the voltage drop. This assumes, of course, that one can guess or motivate the main features in advance. For instance, for conductive molecular devices, the bias voltage will drop mainly at the left and right contact regions if the contacts have low transparency or over the whole molecular region if the contacts are very transparent.  

In our approach, we introduce new parts on the left ($C_{L'}$) and right ($C_{R'}$) within the region $C$ which extend from their respective leads to the molecular contacts, as shown in Fig. 1.  We will denote by $X$ the interface between the molecule $C_{C'}$ and the regions $C_{L'}$ plus $C_{R'}$. If the voltage drop around the left (right) contact is $\alpha V_b$ ($(1-\alpha) V_b$), then the potential shifts are applied in the following way (which we call the $\Delta\mathbf{H}1$ treatment):
\begin{eqnarray}
\mathbf{H}_{CC}^{'}&=&\mathbf{H}_{CC}+\left(\alpha-\frac{1}{2}\right)eV_b \mathbf{S}_{CC}, \\
\left[\Delta\mathbf{H}_{C_L'}\right]_{\mu\nu}&=&-e \alpha V_b \left[
  \mathbf{S}_{CC}\right]_{\mu\nu}, \label{equ_shl'} 
  \mu \textrm{\bf \;or } \nu \in C_{L'}, \\
\left[\Delta\mathbf{H}_{C_R'}\right]_{\mu\nu}&=&e (1-\alpha) V_b \left[
  \mathbf{S}_{CC}\right]_{\mu\nu}, 
  \mu \textrm{\bf \;or } \nu \in C_{R'}. \label{equ_shr'} 
\end{eqnarray}
Because the potential shift is applied to a matrix element when \textit{either} orbital index is in the $C_{L',R'}$ part, the voltage drop will be slightly smeared around the interface $X$.  To explicitly show the role of $\mathbf{S}_{CC}$ in (\ref{equ_shl'}) and (\ref{equ_shr'}), we also drop the potential using (called the $\Delta\mathbf{H}0$ treatment)
\begin{eqnarray}
\left[\Delta\mathbf{H}_{C_L'}\right]_{\mu\nu}&=&-e \alpha V_b \left[
  \mathbf{S}_{CC}\right]_{\mu\nu}, \label{equ_shl''}
  \mu, \nu \in C_{L'}, \\
\left[\Delta\mathbf{H}_{C_R'}\right]_{\mu\nu}&=&e (1-\alpha) V_b \left[
  \mathbf{S}_{CC}\right]_{\mu\nu},
  \mu, \nu \in C_{R'}. \label{equ_shr''}
\end{eqnarray}
In either case, $\mathbf{G}_{CC}(E)$ is determined by an equation analogous to Eq. (\ref{GccVb}) in the SC case,
%\begin{widetext}
\begin{eqnarray}
\label{equ_Gc''}
\mathbf{G}_{CC}(E)&=&\left(E\mathbf{S}_{CC}-\left[\mathbf{H}_{CC}^{'}+\Delta\mathbf{H}_{C_L'}+
                     \Delta\mathbf{H}_{C_R'} \right. \right. \\ \nonumber 
                 &&+\mathbf{\Sigma}_L(E+\frac{eV_b}{2}) \left. \left.
                   +\mathbf{\Sigma}_R(E-\frac{eV_b}{2})\right]\right)^{-1} \,.
\end{eqnarray}
%\end{widetext}
The initial $\mathbf{H}_{CC}$ matrix here comes from a separate DFT calculation using a large $L$-$C$-$R$ supercell. Finally, the current is calculated as usual through Eqs. (\ref{equ_t'}) and (\ref{equ_I}).

So far we have considered the simplest case of low transparency contacts so that the voltage 
drops in only two places near the contacts.  In this simplest case the $\alpha$ parameter here 
has the same role as the $\eta$ parameter in Ref. \onlinecite{datta0}.
For a real device, the voltage drop may be much more complicated, 
and there may be several different voltage drops inside the device region. 
However, in our method all the factors affecting the voltage drop have been taken into account
at the DFT level, and it is straightforward 
to generalize the present non-SC approach for these more complicated situations (see calculations 
for system B in Section IV) as long as the main features of the voltage drop are known.  
Actually, we will show later that results of this non-SC approach are not very sensitive to the 
choice of the voltage drop.  If we assume that the form of the drop is not affected significantly 
by a change in the bias voltage itself, then the main features of the voltage drop in a system can
be determined by a single self-consistent calculation using a relatively small bias voltage.  
%Or more simply, we can guess the main features and then check by simply comparing the 
%transmission functions from the SC and non-SC calculations.  

%------------------------------ Fig. 3 -------------------------------------
\begin{figure}[tb] \label{fig_cells}
\includegraphics[angle=0,width= 8.5cm]{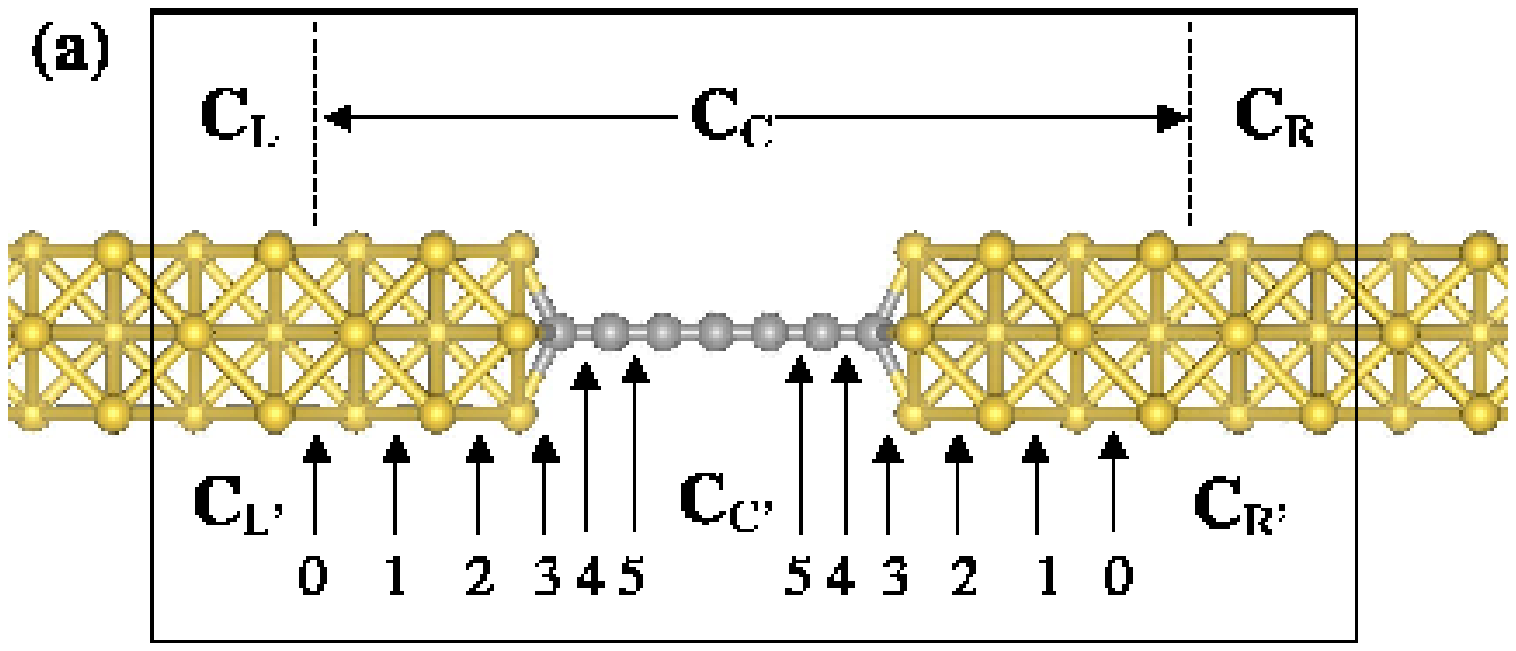}
\includegraphics[angle=0,width= 8.5cm]{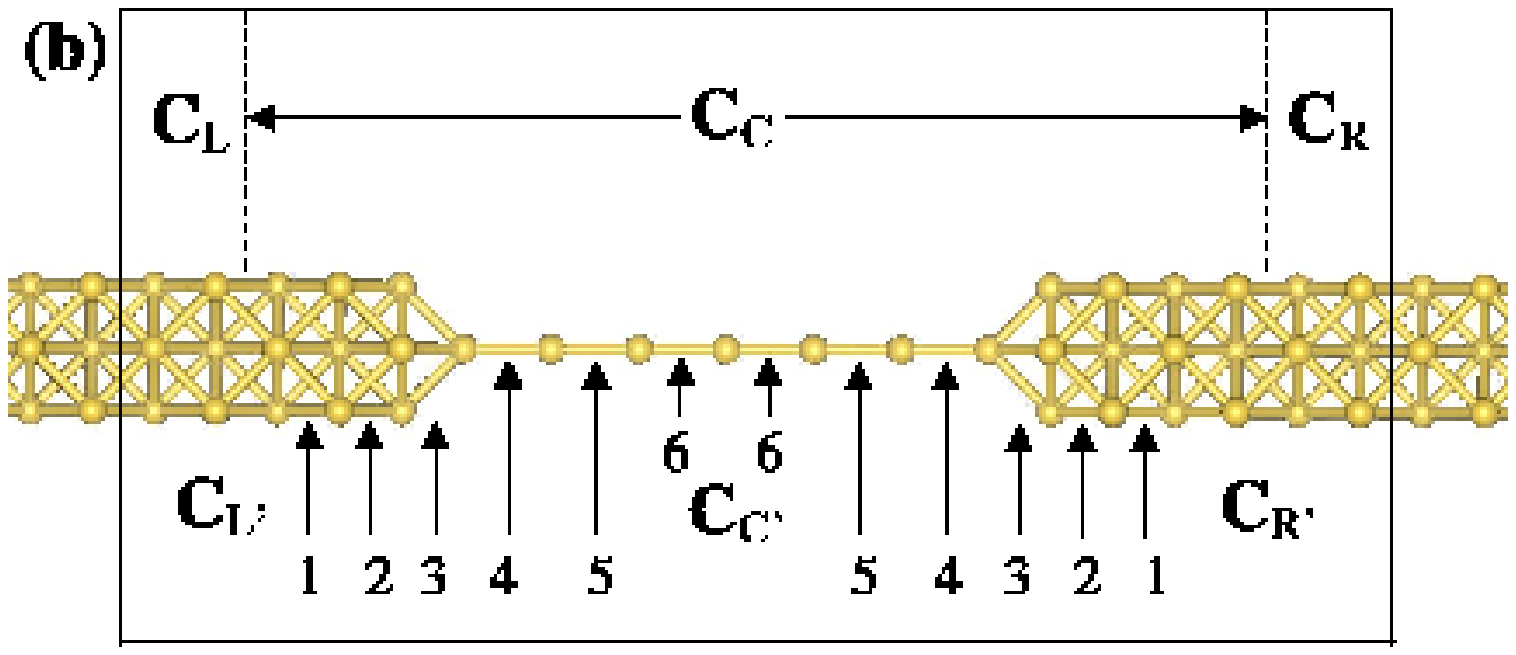}
\caption{(Color online) Systems calculated: (a) a chain of 7 carbon atoms sandwiched between two Al leads in the (001) direction of bulk Al, (b) a chain of 7 Al atoms sandwiched between the same leads.  The C-Al distance in (a) is 1{\AA} and the C-C distance is 1.323{\AA}.  The Al-Al (chain) distance in (b) is 2.86{\AA} and the Al-Al(surface) distance is 2.025{\AA} (i.e., the two Al atoms at the ends of the chain are in their bulk positions).  The notations for different parts are the same as in Fig. 1.  The numbers 0, 1, 2, 3, 4, 5, 6 denote different interfaces, called the interface $X$, between $C_{L'}$ or $C_{R'}$ and $C_{C'}$ which are used in the present non-self-consistent approach.  }
\end{figure}

\section{Applications: Chains of Carbon or Aluminum} 

We report calculations of $I$-$V$ curves for two systems: a \textit{carbon chain (system A)} or an \textit{aluminum chain (system B)} sandwiched between two aluminum leads in the (001) direction of bulk Al.  The structures are shown in Fig. 3. No further atomic relaxation is performed for 
simplicity and for direct comparison with previous results. 

Our implementation of the transmission calculation is independent of the DFT part.  Therefore, it can be easily combined with any DFT package that uses a localized basis set.  As an application, here we combine it with the very efficient full DFT package SIESTA,\cite{siesta} which adopts a LCAO-like and finite-range numerical basis set and makes use of pseudopotentials for atomic cores.  
In our calculations we adopt a single zeta (SZ) basis set. To check the 
convergence of the results, we also calculate the equilibrium transmission function using a single zeta plus polarization (SZP) basis set.  The difference is found to be minor. The PBE version of the generalized gradient approximation \cite{pbe} is adopted for the electron exchange and correlation, and optimized Troullier-Martins pseudopotentials \cite{tmpp} are used for the atomic cores.  The initial density matrix of the region $C$ is obtained from a separate DFT calculation using a large $L$-$C$-$R$ supercell.

%------------------------------ Fig. 4 ---------------------------------------
\begin{figure}[tb] \label{fig_vdmap}
\includegraphics[angle=-90,width= 5.0cm]{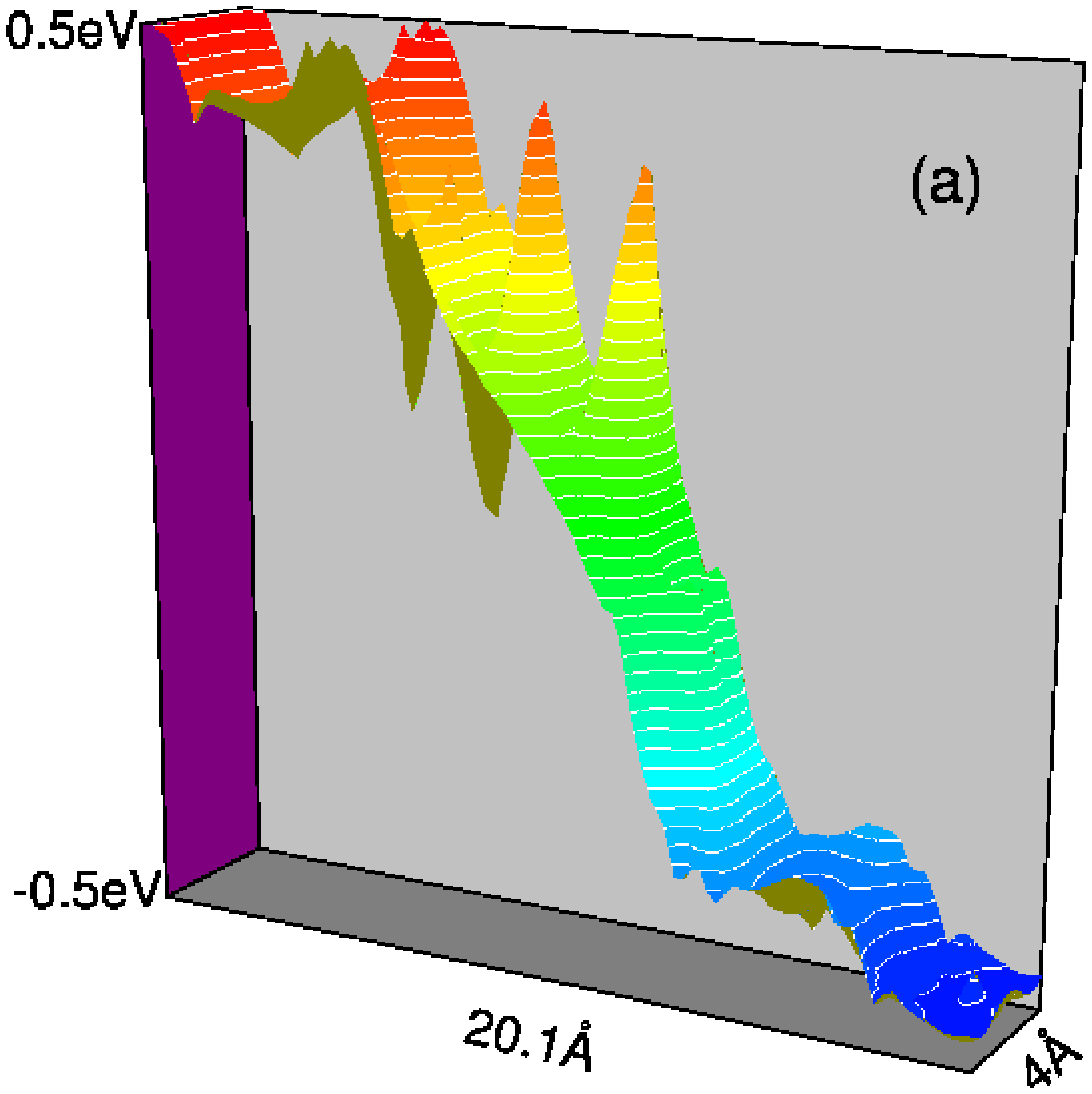}
\includegraphics[angle=-90,width= 5.0cm]{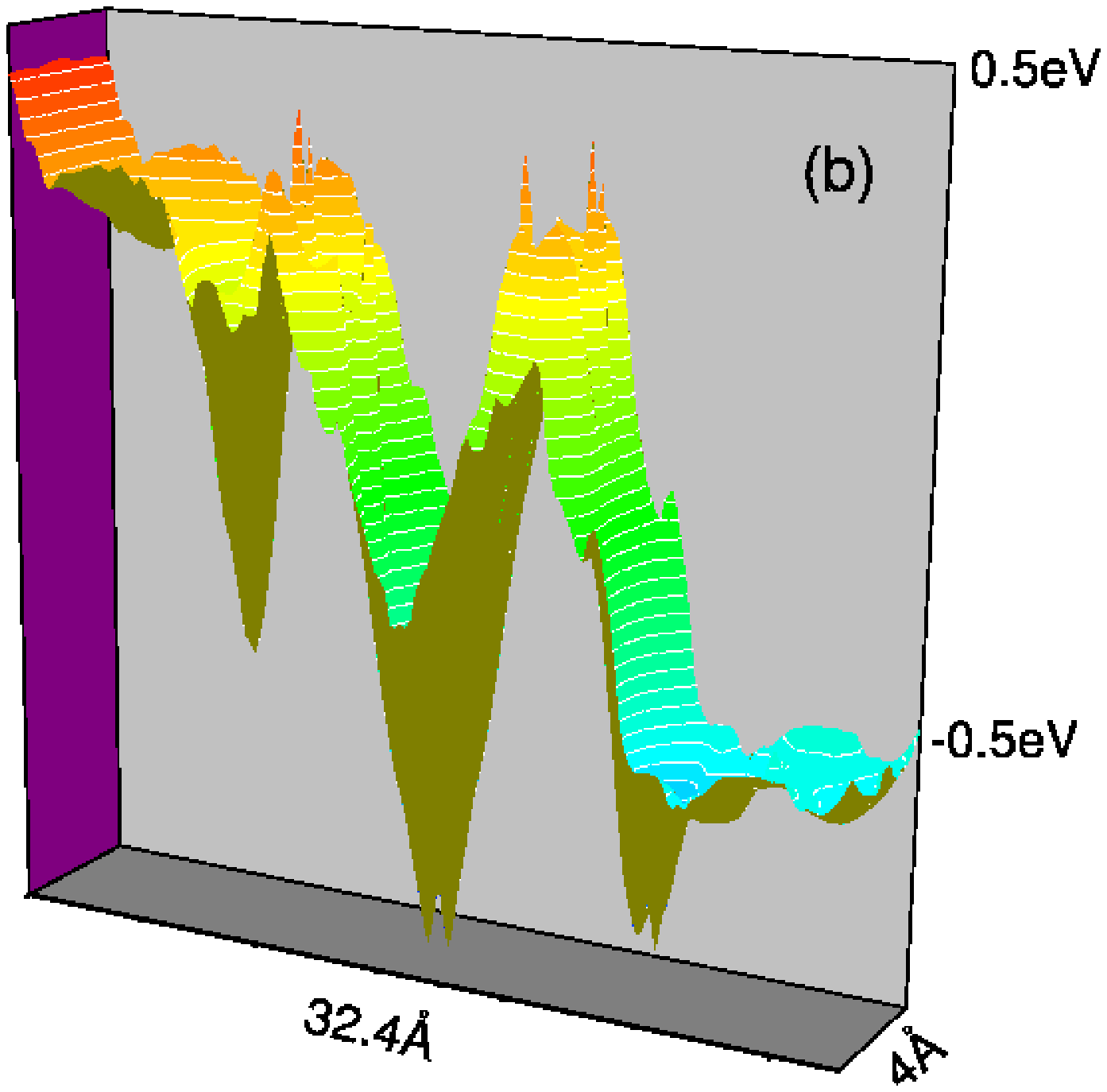}
\caption{(Color online) Voltage drop in a plane going through the atomic chain for the $C_C$ region of (a) the C chain and (b) the Al chain, as shown in Fig. 3, for an applied bias of  -1 V. Note that (1) in (a) the voltage drop mainly occurs around the right contact region while in (b) it occurs over the entire chain, and (2) the oscillation in (b) is much larger than that in (a). This is because the electrons in the Al chain are more free than that in the C chain; as a result, the polarization induced by the bias is larger in the Al chain. 
}
\end{figure}

There are two main reasons for us to choose to study these systems.  First, the transmission function $T(E)$ of system A has been calculated by both the TranSiesta \cite{transiesta} and MCDCAL \cite{mcdcal_chain} packages using a SZ basis set.  So we can make a direct comparison to previous results.  Second, systems A and B typify two different voltage drop behaviors (although both the C and Al chains are conductive).  In system A, because the molecule-lead contact is a hetero-interface, the voltage drop will mainly occur around the two interfaces [see Fig. 4 (a), note that the voltage drops around the two interfaces are actually asymmetric].  In contrast, in system B the molecule-lead contact is a homo-interface, and furthermore the two Al atoms at the ends of the chain are at their bulk positions. So the voltage drop will occur over the entire Al chain in some way [see Fig. 4 (b)].  Our purpose is to see whether our non-SC approach can handle these different conditions.

%----------------------- Fig. 5 ----------------------------------------
\begin{figure}[bt] \label{fig_t0_pdos_a_b}
\includegraphics[angle=0,width=8.5cm]{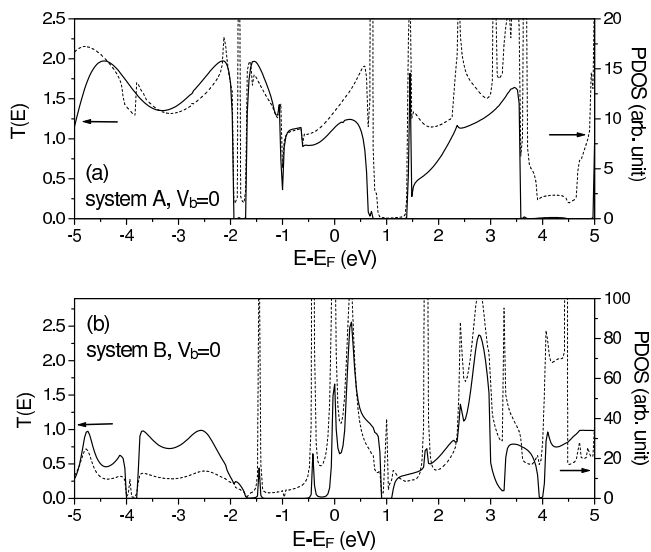}
\caption{Calculated transmission function $T(E)$ (solid line) and projected density 
of states (PDOS, projected on the chains, dashed line) 
by the self-consistent approach for (a) the carbon chain (system A) and
(b) the aluminum chain (system B) under zero bias voltage. When $T(E)$ is significantly different from the PDOS, it means that a localized state exists at that energy. }
\end{figure}

\subsection{Transmission functions}

In Fig. 5 we show the calculated transmission functions and PDOS (projected on the chains) for system A and B under zero bias voltage. As it can be seen, the transmission function generally follows the PDOS except for some localized states (for instance, around 4 $\sim$ 5 eV in (a) and 1 eV in (b)) which are not coupled with the left or the right lead. Our result of $T(E)$ for system A is in very good agreement with the previous results from TranSiesta and MCDCAL packages \cite{transiesta,mcdcal_chain} (see Fig. 6 (a) of Ref. \onlinecite{transiesta}).

%----------------------------- Fig. 6-------------------------------------
\begin{figure}[tb] \label{fig_t_c7al_1v}
\includegraphics[angle=0,width=8.5cm]{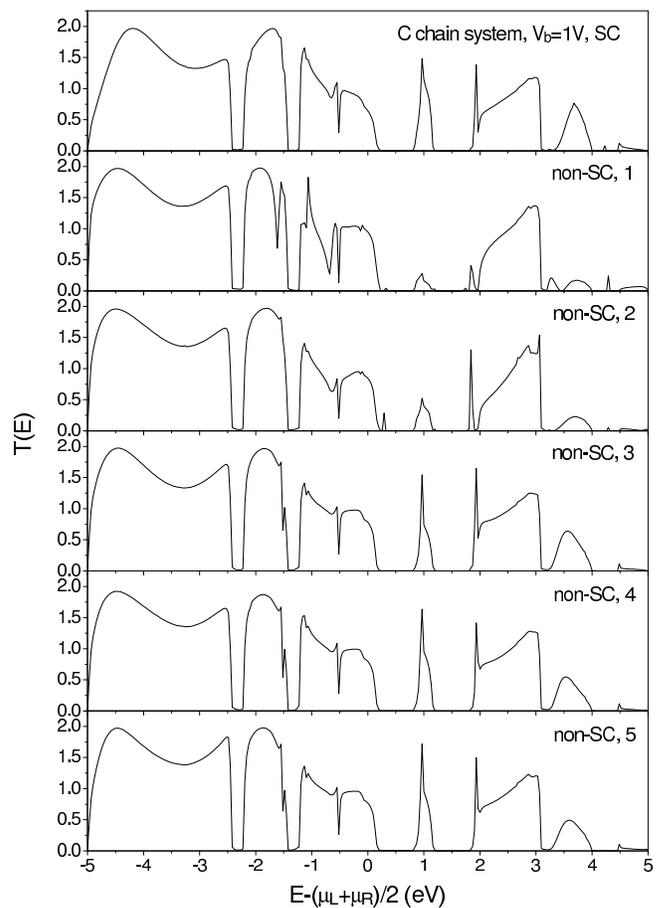}
\caption{Transmission function $T(E)$ for the carbon chain (system A) under
1 V bias ($\alpha=0.5$) from the SC approach and the non-SC  
approach ($\Delta\mathbf{H}1$ treatment) with different choices 
for the interface $X$ between $C_{C'}$ and the $C_{L'}$ or
$C_{R'}$ parts (denoted by `1', ..., `5')
as shown in Fig. 3 (a). Note the similarity between the different non-SC results and the fully SC result once $X$ is placed in the contact regions ($\ge 3$). }
\end{figure}

In Fig. 6 we show, for system A under a bias of 1.0V ($\alpha=0.5$), the calculated $T(E)$ by the SC approach and the non-SC approach ($\Delta\mathbf{H}1$ treatment) with different choices for the interface $X$ between the $C_{L'}$ or $C_{R'}$ and $C_{C'}$ parts.  The first thing we note is that our SC result for 1V bias is in very good agreement with the previous self-consistent results \cite{transiesta,mcdcal_chain} (see Fig. 6 of Ref. \onlinecite{transiesta}).  When we change the interface $X$ (denoted by the different numbers in Fig. 6) from deep in the leads to the contact regions (i.e., interface $X$ = 1 $\rightarrow$ 2 $\rightarrow$ 3), the non-SC result varies significantly. However, as $X$ moves into the contact regions (i.e., interface $X$ = 3, or 4, or 5), the result becomes very close to the SC result and insensitive to the exact position of $X$.  This result is just what we expect because in system A the bias voltage drops mainly around the hetero-interface contact regions. This can be regarded as an advantage of the present non-SC approach:its result is not strongly dependent on the technical choice.

%-------------------------------- Fig. 7 ----------------------------------
\begin{figure}[tb] \label{fig_t_al7al_1v}
\includegraphics[angle=0,width=8.5cm]{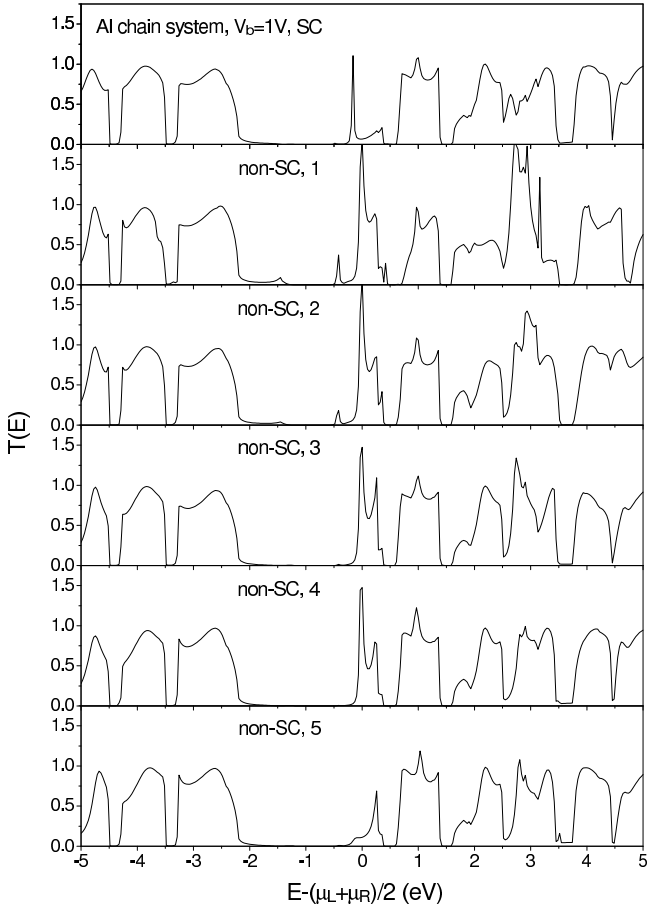}
\caption{Calculated transmission function $T(E)$ for the Al chain system under
1 V bias ($\alpha=0.5$, $\Delta\mathbf{H}1$ treatment for non-SC) by the SC and non-SC approaches.
%: present self-consistent (SC) approach,
%existing non-self-consistent (non-SC) approach, and present non-self-consistent
%approach with different chooses of the interface between the $C_{L'}$ or
%$C_{R'}$ and $C_{C'}$ regions (denoted by `1', `2', `3', `4', and `5')
%as shown in Fig. \ref{fig_cells} (b).
The notations are similar to those in Fig. 6}
\end{figure}

Similar calculations of $T(E)$ for system B under $V_b$ = 1.0V ($\alpha=0.5$, $\Delta\mathbf{H}1$ treatment for non-SC) are shown in Fig. 7.  Again, after the interface $X$ is moved into the contact regions the different non-SC result become quite close.  However, compared to the case of system A, the agreement with the SC result is not good, especially in the energy range around the (averaged) Fermi level: the transmission from the non-SC calculations (except for interface $X$ = 5) is noticeably larger than that from the SC calculation. This substantial disagreement originates in the difference between the self-consistent effective potential [Fig. 4(b)] and the non-self-consistent one assumed in the non-SC approach.

%-------------------------------- Fig. 8 -------------------------------------
\begin{figure}[tb] \label{fig_didv_a_vd}
\includegraphics[angle=0,width=8.5cm]{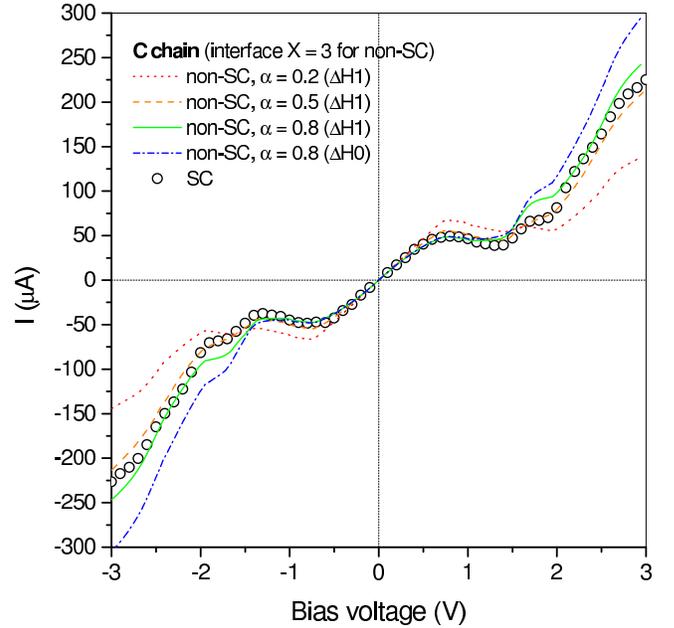}
\caption{(Color online) Calculated $I$-$V$ curves for the C chain system by the SC approach and the non-SC approach with interface $X$ = 3 [as shown in Fig. 3 (a)].  Three different voltage drops are considered: $\alpha$ = 0.2, 0.5, 0.8.  The $\Delta\mathbf{H}0$ and $\Delta\mathbf{H}1$ treatments are adopted for $\alpha$ = 0.8. The simple non-SC approach does very well using $\alpha=0.5$ or $0.8$ and the $\Delta\mathbf{H}1$ treatment.
}
\end{figure}

\subsection{$I$-$V$ curve of system A}
%effect of different voltage drops, $\Delta\mathbf{H}0$ and $\Delta\mathbf{H}1$ treatments}

In order to show effects of different voltage drops and the difference between the $\Delta\mathbf{H}0$ and $\Delta\mathbf{H}1$ treatments [see Eqs. (\ref{equ_shl'})-(\ref{equ_shr''})] for potential shifting, we give in Fig. 8 the calculated $I$-$V$ curves for system A from the non-SC approach with interface $X$ = 3 compared to the SC result. We do the non-SC calculations for three different voltage drops ($\alpha$ = 0.2, 0.5, 0.8) for the $\Delta\mathbf{H}1$ treatment.  In addition, for $\alpha$ = 0.8 we do the non-SC calculation with the $\Delta\mathbf{H}0$ treatment.  Among the three different voltage drops, the result for $\alpha$ = 0.2  is in poor agreement with the SC result while those for $\alpha$ = 0.5 and 0.8 are in good agreement.  This makes sense in view of the main features of the voltage drop in Fig. 4(a): the bias voltage will mainly drop around the left (right) contact for a positive (negative) bias.  However, the small difference in $I$-$V$ curve between $\alpha$ = 0.5 and 0.8 indicates that the $I$-$V$ characteristics is actually not sensitive to the exact change in voltage drop.  By comparing the SC result and the two non-SC results for $\alpha$ = 0.8, we see that the $\Delta\mathbf{H}1$ treatment improves the result markedly. This is understandable because voltage drops in real physical systems are not sharp step-functions but are somewhat smeared out.

%-------------------------------- Fig. 9 -------------------------------------
\begin{figure}[b] \label{fig_didv_a}
\includegraphics[angle=0,width=8.5cm]{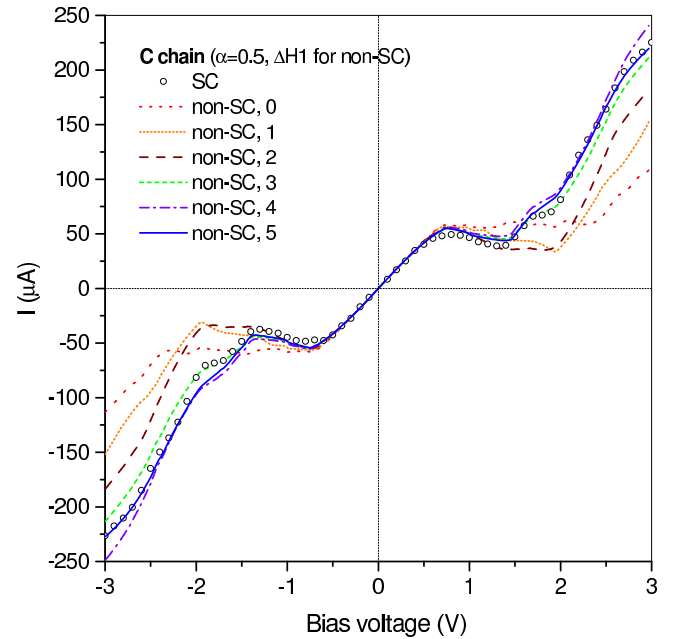}
\caption{(Color online) Calculated $I$-$V$ curves for the C chain system by the SC approach and the non-SC approach with different interface $X$ indicated by the numbers. $\alpha$ = 0.5 and $\Delta\mathbf{H}1$ treatment are adopted for the non-SC calculations.
%the non-self-consistent approach in recent literatures; the numbers %`1', `2', `3', `4', `5' indicate the results from the present %non-self-consistent approach using different interfaces between %the $C{L'}$ or $C_{R'}$ and $C_{C'}$ regions, as shown in Fig. \ref{fig_cells} (a).
}
\end{figure}

%--------------------------------- Fig. 10 --------------------------------
\begin{figure}[tb] \label{fig_didv_b}
\includegraphics[angle=0,width=8.5cm]{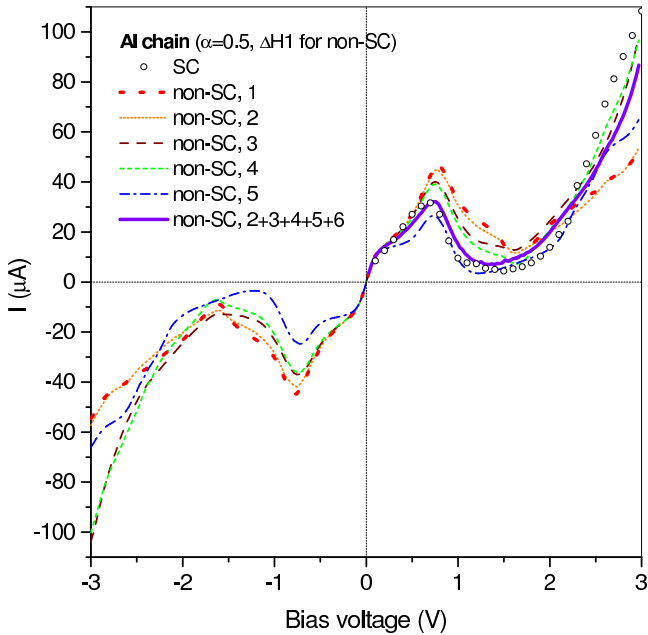}
\caption{(Color online) Calculated $I$-$V$ curves for the Al chain system by the SC and non-SC approaches.  The notations are similar to those in Fig. 9.  `2+3+4+5+6' means a combined interface $X$ (see Fig. 3 (b) ) and each interface bears 1/10 of a bias voltage (see Section III C).}
\end{figure}

\subsection{$I$-$V$ curves of systems A and B: comparison of different approaches}

$I$-$V$ curves for systems A and B calculated by different approaches are given in Figs. 9 and 10, respectively. For the non-SC approach, $\alpha$ = 0.5 and the $\Delta\mathbf{H}1$ treatment are always adopted. It turns out that for small bias voltages ($V_b$ $<$ 0.3V) all the different treatments (the SC, the non-SC with different interface $X$), all give similar results, i.e., the effect of the different voltage drops is very small. Along with the increase of bias voltage, the difference among the different calculations becomes more and more significant. 

For the carbon chain (system A), because the molecule-electrode contact is a hetero-interface and therefore the voltage drop occurs mainly around the contact regions, the SC result and the non-SC results with the interface $X$ located around the two contact regions are in good agreement. As for the transmission function in Fig. 6, once the interface $X$ is within the contact regions the result is quite insensitive to the technical choice. This indicates that for systems made of conductive molecular junctions coupled with metallic electrodes through hetero-interfaces, the present non-SC approach works quite well and can give nearly \textit{quantitatively} correct answers.

For the aluminum chain (system B) the molecule-electrode contact is a homo-interface and the two Al atoms at the ends of the chain are at their bulk positions; therefore, the voltage drop is not localized around the contact regions. Consequently, the result from the non-SC approach with the interface $X$ located around the contact region is not in good agreement with the SC result.  In order to further verify our analysis for system B, we generalize the present non-SC approach for a voltage drip occurring at multiple points: We use a combined interface $X$ = 2+3+4+5+6 in which each layer bears a voltage drop of $V_b$/10. Because of the role played by the overlap matrix in Eqs. (\ref{equ_shl'}) and (\ref{equ_shr'}), the resulting voltage drop will occur over the entire device region. The calculated $I$-$V$ curve by the generalized non-SC approach is given in Fig. 10 by a violet solid line.  The overall agreement with the SC result is remarkably improved, indicating that our analysis is reasonable.

\subsection{Limitation of the present approaches}

For finishing the discussion we would like to point out the cases where the present method
will not work. Obviously, the present method is only valid for steady state coherent electron 
transport through metal-molecule-metal systems; there are basically two cases
where our method does not work: (1) electron transport in Coulomb
blockade regime, for both the
SC and non-SC approaches, and (2) cases where the main feature of voltage drop
is sensitive to the value of the bias voltage itself, for the non-SC approach.
In the first case, the contact barrier is so high that the molecule
and the leads are essentially separated, and as a result, the molecular
chemical potential is generally different from the Fermi energies of the
leads even under zero bias.
Because in our DFT+NEGF approach there is only
one Fermi energy under zero bias, it will fail in this case.
The second case is just the opposite to that assumed in our
non-SC approach. We don't know at this moment what systems will have this
behavior or whether such kind of systems exist. But this can be easily checked
by doing selfconsistent calculations for several different bias voltages
within the bias range interested.

\section{Summary}

A full self-consistent DFT-electronic-structure-based Green function method has been proposed and implemented for electron transport from molecular devices.  Our method is simple and straightforward while strict.  The implementation is very independent of the DFT electronic structure part; it can be easily combined with any electronic structure package using a localized basis set.  In an effort to avoid the extremely burdensome computational cost for large systems or for $I$-$V$ characteristic analysis, we developed an approximate non-self-consistent approach in which the change in effective potential caused by a bias in the device region of a system is approximated by the main features of the voltage drop.  

As applications of our methods, we calculated the $I$-$V$ curves for two different systems with 
different typical voltage drops: a carbon chain and a aluminum chain sandwiched between two 
aluminum electrodes.  Our self-consistent results are in very good agreement with those from 
other calculations. For both systems the present non-SC approach can give results in good 
agreement with the self-consistent results, indicating that it is a good approximate method 
with high efficiency for $I$-$V$ characteristic analysis \cite{tddft}
(more then one order of magnitude 
faster for moderate systems). It is straightforward to generalize this non-SC approach to 
deal with any kind of voltage drop situation.

%Our numerical calculations have shown that, overall, the $I$-$V$ characteristics %of a molecular device is actually a quite insensitive quantity to the %change of voltage drop inside the device region. 

%\acknowledgments
This work was supported in part by the NSF (DMR-0103003).


\begin{thebibliography}{99}
\bibitem{test}      For instance, our test calculation for benzene molecule 
                    sandwiched between two Au electrodes
                    showed that the molecule-electrode charge transfer depends significantly
                    on the adsorption site of the molecule on the Au surface.
\end{thebibliography}
\end{document}